\documentclass{PoS}
\usepackage{amsmath,amssymb,amsfonts}  
\usepackage{braket}
\newcounter{MBQ}

\newcommand{\be}{\begin{equation}}
\newcommand{\ee}{\end{equation}}
\newcommand{\bea}{\begin{eqnarray}}
\newcommand{\eea}{\end{eqnarray}}
\newcommand{\bi}{\begin{itemize}}
\newcommand{\ei}{\end{itemize}}
\newcommand{\ben}{\begin{enumerate}}
\newcommand{\een}{\end{enumerate}}
\newcommand{\bt}{\begin{tabular}}
\newcommand{\et}{\end{tabular}}

\newcommand{\nn}{\nonumber}

\newcommand{\nnm}[1]{n_{#1-}}
\newcommand{\nnp}[1]{n_{#1+}}

\newcommand{\als}{\alpha_s}

\global\long\def\order#1{\mathcal{O}\left(#1\right)}

\allowdisplaybreaks

\title{Subleading-power 
${N}$-jet operators and the LBK amplitude in SCET}

\ShortTitle{Subleading-power 
${N}$-jet operators and the LBK amplitude in SCET}

\author{\speaker{Martin Beneke}\thanks{TUM-HEP-1120/17}, Mathias Garny, 
Robert Szafron, Jian Wang\\
        Physik Department T31, James-Franck-Stra\ss{}e~1, Technische Universit\"at M\"unchen, \\
        85748 Garching, Germany}

\abstract{We construct a complete and minimal basis of subleading-power 
${N}$-jet operators in position-space soft-collinear effective theory (SCET)
and discuss how the Low-Burnett-Kroll amplitude is recovered in this 
framework. We begin a systematic investigation of the anomalous dimension of 
next-to-leading power $N$-jet operators in view of resummation of 
logarithmically enhanced terms in partonic cross sections and discuss the  
explicit result at the one-loop order for fermion-number two $N$-jet 
operators at the second order in 
the power expansion parameter of SCET.}

\FullConference{13th International Symposium on Radiative Corrections 
(Applications of Quantum Field Theory to Phenomenology)\\
                 25-29 September, 2017 \\
                 St. Gilgen, Austria}

\begin{document}

\section{Introduction\label{sec:intro}}

\noindent 
In this proceedings contribution, we summarize results from our study of 
subleading-power effects in high-energy scattering, in particular the 
anomalous dimension of subleading-power $N$-jet 
operators~\cite{Beneke:2017ztn} in the framework 
of position-space soft-collinear effective theory (SCET). An $N$-jet operator 
describes the hard scattering of $N$ energetic, massless particles with 
(outgoing) momenta $p_i$ with $s_{ij}=2 p_i\cdot p_j+i0$, $i,j=1\ldots N$, 
and all $s_{ij}$ of the order of the square of some large scale $Q$. The 
ultraviolet (UV) divergences of such SCET operators are related to the 
universal infrared (IR) properties of soft and collinear 
radiation~\cite{Becher:2009cu}. Up to the two-loop order the 
leading-power soft-collinear anomalous dimension has the very 
simple structure 
\begin{equation}
\Gamma = -\gamma_{\rm cusp}(\alpha_s) \sum_{i<j} 
\mathbf{T}_i\cdot\mathbf{T}_j \ln\left(\frac{-s_{ij}}{\mu^2}\right) 
+ \sum_{i}
\gamma_i(\alpha_s) 
\label{eq:LPanomalousdim}
\end{equation}
in colour-operator notation. Given the advances in the understanding 
of multi-loop corrections to the leading-power anomalous dimension 
\cite{Almelid:2015jia, Moch:2005id}, 
it is also timely to ask about the next-to-leading power (NLP) term in 
the $M/Q$ expansion, where $M\ll Q$ refers to a smaller scale generated 
by soft or collinear radiation. Besides shedding light on the formal 
structure of IR divergences at subleading power, the NLP $N$-jet anomalous 
dimension is a key ingredient for summing logarithmically enhanced loop 
effects to all orders in perturbation theory. NLP logarithms have 
recently been investigated at fixed NNLO accuracy for the threshold limit 
of the Drell-Yan process~\cite{Bonocore:2016awd} 
and 0-jettiness observables \cite{Moult:2016fqy,Boughezal:2016zws}. 


\section{SCET essentials}

\noindent
We consider physical processes for which the virtuality of 
collinear modes in any of the $N$ jet directions is of the same order, 
and parametrically larger than the one of the soft mode. The power-counting 
parameter $\lambda$ is set by the transverse momentum $p_{\perp i} \sim 
\lambda$ of collinear momenta with virtuality ${\cal O}(\lambda^2)$.
The components of soft momentum are all ${\cal O}(\lambda^2)$ and 
consequently soft virtuality scales as $\lambda^4$. The term 
``NLP'' refers to ${\cal O}(\lambda)$ and ${\cal O}(\lambda^2)$, since 
the first non-vanishing power correction to most physical processes of 
interest is ${\cal O}(\lambda^2)$.

After integrating out the hard interactions, the infrared physics 
is described by the SCET Lagrangian including all subleading-power 
interactions. For $N$ widely separated collinear directions, the Lagrangian 
\begin{equation}
{\cal L}_{\rm SCET} = \sum_{i=1}^N {\cal L}_i(\psi_i,\psi_s) 
+{\cal L}_s(\psi_s)
\label{eq:SCETI}
\end{equation}
is the sum of $N$ copies of collinear Lagrangians with $N$ pairs of separate 
light-like reference vectors $n_{i \pm }$, $i=1,\dots,N$ satisfying 
$\nnm{i}\cdot\nnm{j} = {\cal O}(1)$. The collinear fields $\psi_i$ all 
interact with the same soft field $\psi_s$ but not among each other. 
The following properties of the position-space SCET formulation 
\cite{Beneke:2002ph,Beneke:2002ni} are essential for the following analysis:
\begin{itemize}
\item In products with collinear fields $\psi_i(x)$, soft 
fields $\psi_s(x)$ must be multipole-expanded in $x$ around  
$x_{i-}^\mu = (\nnp{i}x)\,\nnm{i}^\mu/2$. 
\item There are no purely collinear subleading-power interactions.  
That is, writing ${\cal L}_i(\psi_i,\psi_s) = \sum_{n=0}^\infty 
{\cal L}_i^{(n)} (\psi_i,\psi_s)$ any interaction vertex from the subleading 
power Lagrangians $n> 0$ contains at least one soft field. 
The NLP Lagrangians $n=1,2$ are known 
and higher-order terms can be easily constructed.
\item The SCET Lagrangian is invariant under $N$ separate collinear 
gauge transformations, which operate on a single collinear direction, 
and a soft gauge transformation under which all fields transform, 
see Ref.~\cite{Beneke:2002ni}.
\end{itemize}

\section{\boldmath Subleading-power $N$-jet operator basis}
\label{sec:basis}

\noindent
We construct a complete and minimal basis of subleading 
$N$-jet operators in SCET. The general structure of an $N$-jet operator 
\be
J = \int dt\; C(\{t_{i_k}\}) \,J_s(0) 
\prod_{i=1}^N J_i(t_{i_1},t_{i_2},\dots)
\label{eq:Njetop}
\ee
can be described by products of operators $J_i$ associated with collinear 
directions $\nnm{i}$,  
each of which is itself composed of a product of $n_i$ gauge-invariant 
collinear ``building blocks'' $\psi_{i_k}$,
\be
J_i(t_{i_1},t_{i_2},\dots) = \prod_{k=1}^{n_i} \psi_{i_k}(t_{i_k}\nnp{i})\,,
\ee
and a purely soft-field operator $J_s$. 
In general, each of the collinear building blocks 
is integrated over the corresponding light-like direction in position space,
$C(\{t_{i_k}\})$ is a hard matching coefficient, and $dt=\prod_{ik} dt_{i_k}$.
Apart from the light-like displacements, the 
operators are evaluated at position $X=0$, corresponding to the location of 
the hard interaction.

The guiding principle for constructing building blocks is the requirement of 
collinear and soft gauge covariance. Because each collinear sector transforms 
under its own collinear gauge transformation, each collinear building block 
must be a collinear gauge singlet. However, the soft field may interact with 
different collinear sectors so we only need to assume that collinear building 
blocks transform covariantly under the soft gauge transformation, hence 
\be
J_i(x) \xrightarrow{\textrm{coll.}} J_i(x),\qquad
J_i(x) \xrightarrow{\textrm{soft}} U_s(x_{i-}) J_i(x)\,, 
\ee
where $U_s$ must be taken in the 
(not necessarily irreducible) colour representation of $J_i$. For the 
matrix adjoint representation we would have $J_i(x) 
\xrightarrow{\text{soft}} U_s(x_{i -}) J_i(x) U^\dagger_s(x_{i-})$ with 
$U_s$ in the fundamental. 
The elementary collinear-gauge-invariant collinear building blocks are 
given by 
\be
\psi_{i}(t_{i}\nnp{i}) \in 
\left\{ \begin{array}{ll} 
\chi_i(t_{i}\nnp{i}) \equiv W_{i}^\dag\xi_{i} & 
\hspace*{0.5cm} \mbox{collinear quark}\\[0.2cm] 
{\cal A}_{\perp i}^\mu(t_{i}\nnp{i})\equiv 
W_{i}^\dag [iD_{\perp i}^\mu W_{i}] & 
\hspace*{0.5cm} \mbox{collinear gluon} 
\end{array}\right.
\label{eq:elementaryblock}
\ee
for the collinear quark and gluon field in the $i$-th direction, respectively. 
$W_i$ is the path-ordered exponential of $\nnp{i} A_{i}$ 
(``$i$-collinear Wilson line'') and the covariant derivative includes 
only the collinear gluon field. Both, the quark and gluon building blocks 
scale as ${\cal O}(\lambda)$. 
Objects containing $i\nnp{i} D_i$ 
or $i \nnp{i}\partial$ are redundant. At leading power, 
only a single building block contributes to each 
direction, i.e.~$n_i=1$ for all $i=1,\dots,N$,
and the $J_i(t_{i_1},t_{i_2},\dots)$ are given by 
\be
J_i^{A0}(t_{i}) = \psi_{i}(t_{i}\nnp{i})\,.
\ee

\begin{figure}[t]
 \includegraphics[width=0.237\textwidth]{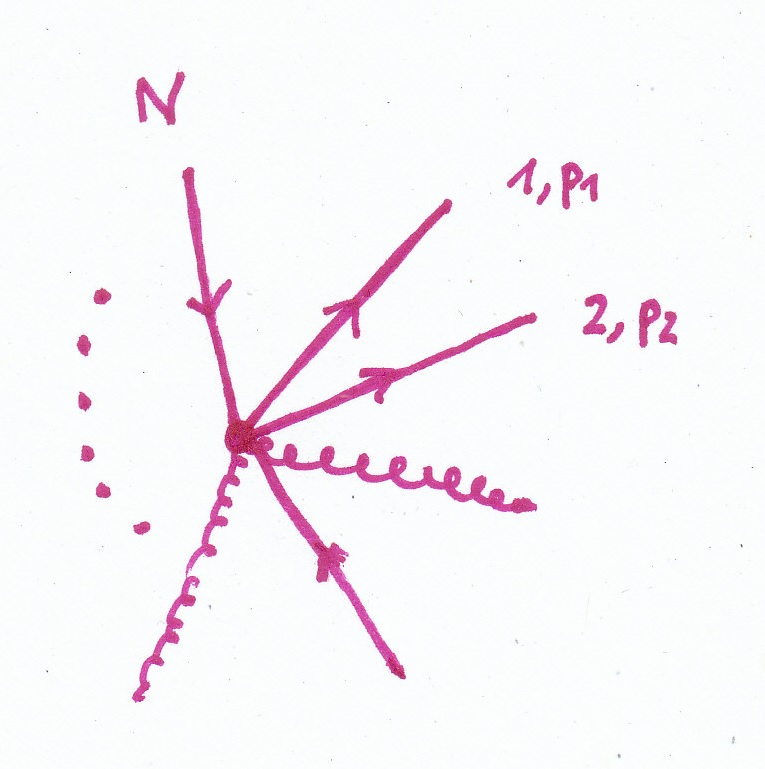}\hspace{0.6cm}
 \includegraphics[width=0.20\textwidth]{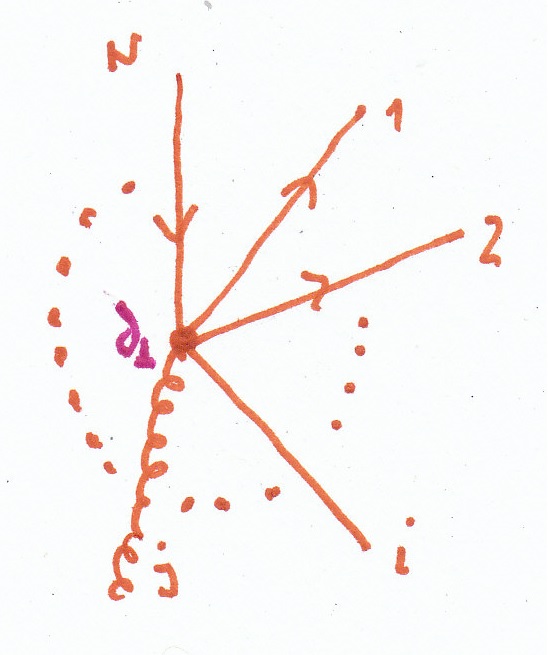}
 \includegraphics[width=0.23\textwidth]{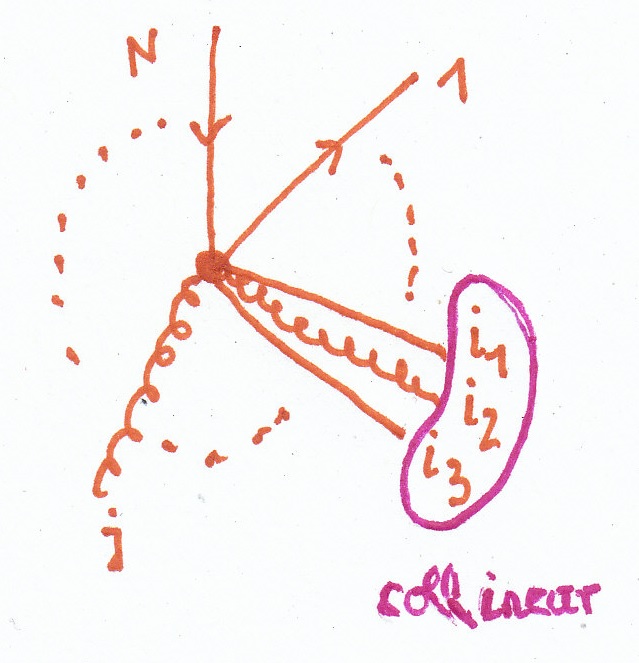}
 \includegraphics[width=0.229\textwidth]{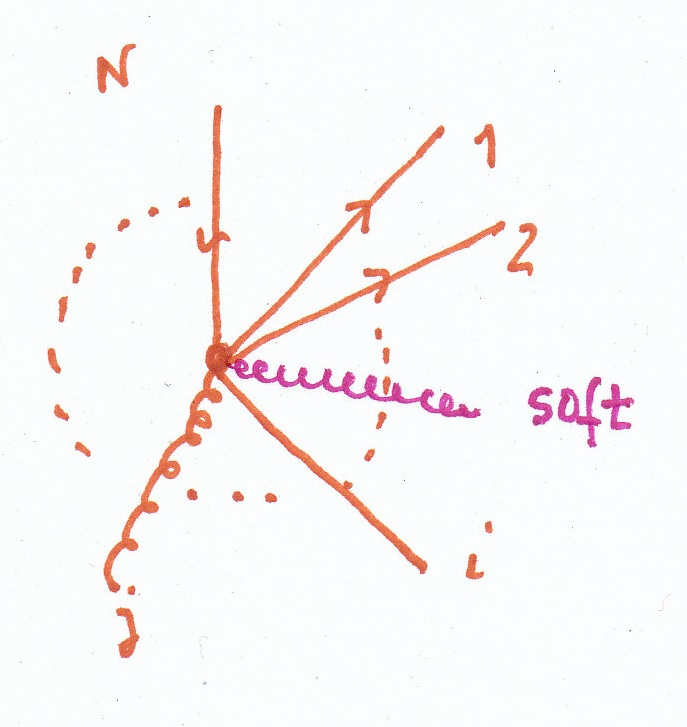}
\caption{Three ways (second to fourth panel) to generate 
subleading-power $N$-jet operators from the leading-power operator 
$\prod_{i=1}^N J_i^{A0}(t_{i})$ (first panel).
\label{fig:njetoperators}}
\end{figure}

We are interested in $N$-jet operators that are suppressed by one or two 
powers of $\lambda$ relative to the leading power. In general, this 
suppression can arise in three ways, shown in Fig.~\ref{fig:njetoperators}: 
\begin{itemize}
\item[(i)] via higher-derivative operators, i.e. acting with either 
$i\partial_{\perp i}^\mu\sim{\cal O}(\lambda)$ or 
$i\nnm{i}D_s\equiv i\nnm{i}\partial+g_s\nnm{i}A_s(x_{i-})
\sim{\cal O}(\lambda^2)$ on the elementary building blocks 
$\psi_{i_k}$. Here it is important to note that since the elementary building 
blocks transform under the soft gauge transformation with $U_s(x_{i-})$, 
the covariant soft derivative is the ordinary derivative for the transverse 
direction and $i\nnm{i}D_s$ for the $\nnm{i}$ projection, which  
guarantees a homogeneous scaling in $\lambda$;
\item[(ii)] by adding more elementary building blocks 
$\chi_i, {\cal A}_{\perp i}^\mu$ in a given direction;
\item[(iii)] via the new elementary collinear building block 
$\nnm{i}\mathcal{A}_{i} \equiv 
W_{i}^\dag i\nnm{i} D_{i} W_{i} -i \nnm{i} D_s$ that appears at 
subleading power, and via purely soft building blocks in $J_s$.
\end{itemize}
In the following, we label operators that consist of a single building block
by $J_i^{An}$, where $n=1,2$ indicates the relative power suppression due to 
additional derivatives. Operators with two, three, ... elementary building 
blocks are denoted as  $J_i^{Bn}, J_i^{Cn}, \ldots$, and here $n$ indicates 
the power suppression relative to $J_i^{A0}$. 
The ${\cal O}(\lambda)$ operator basis consists of 
\bea
J_i^{A1}(t_{i}) &=& i\partial_{\perp i}^\nu J_i^{A0}\,,\\
J_i^{B1}(t_{i_1},t_{i_2}) &=& 
\psi_{i_1}(t_{i_1}\nnp{i})\psi_{i_2}(t_{i_2}\nnp{i})
\in \left\{ \begin{array}{ll}
{\cal A}_{\perp i}^\mu(t_{i_1}\nnp{i}) \chi_i(t_{i_2}\nnp{i}) \\[0.03cm]
\chi_i(t_{i_1}\nnp{i})\chi_i(t_{i_2}\nnp{i}) \\[0.03cm]
{\cal A}_{\perp i}^\mu(t_{i_1}\nnp{i}) 
{\cal A}_{\perp i}^\nu(t_{i_2}\nnp{i}) \\[0.03cm] 
\chi_i(t_{i_1}\nnp{i}) \bar\chi_i(t_{i_2}\nnp{i}) \,,
\end{array}\right.
\eea
and at ${\cal O}(\lambda^2)$
\bea
J_i^{A2}(t_{i}) &=& 
i\partial_{\perp i}^\nu\,i\partial_{\perp i}^\rho J_i^{A0}\,, 
\label{eq:A2}\\[0.1cm]
J_i^{B2}(t_{i_1},t_{i_2})  &\in& \left\{\begin{array}{l} 
\psi_{i_1}(t_{i_1}\nnp{i})i\partial_{\perp i}^\mu\psi_{i_2}(t_{i_2}\nnp{i})
\\[0.1cm]
i\partial_{\perp i}^\mu\big[\psi_{i_1}(t_{i_1}\nnp{i})
\psi_{i_2}(t_{i_2}\nnp{i})\big]\,,
\end{array}\right.
\\
J_i^{C2}(t_{i_1},t_{i_2},t_{i_3}) &=& \psi_{i_1}(t_{i_1}\nnp{i})
\psi_{i_2}(t_{i_2}\nnp{i})\psi_{i_3}(t_{i_3}\nnp{i})\,,
\eea
where 
$\psi_{i_1}\psi_{i_2}$ in $J_i^{B2}$ can be any combination from 
$J_i^{B1}$, and $\psi_{i}$ can be any of the elementary building blocks 
from Eq.~(\ref{eq:elementaryblock}). 
This exhausts the options (i), (ii) from above at 
${\cal O}(\lambda^2)$. Concerning (iii), it can be shown (App.~B of 
Ref.~\cite{Beneke:2017ztn}) that 
$\nnm{i}\mathcal{A}_{i}$ can be eliminated by the collinear field 
equation, and the same is true for the soft covariant derivatives, which   
operate on collinear building blocks in the form
\be
i \nnm{i} D_s \chi_i, \qquad [i \nnm{i} D_s, {\cal A}_{\perp i}^\mu]\,. 
\ee
It follows that one can use a basis 
of collinear building blocks that does not involve soft fields through 
covariant derivatives and is constructed entirely from ordinary transverse 
derivatives and the elementary building block for the quark field 
and the transverse gluon field.

In addition to the collinear building blocks, the $N$-jet operator may also 
contain a pure soft building block $J_s(x)$. In the pure soft sector there is 
no need to perform the SCET multipole expansion of the soft fields and 
therefore the soft gauge transformation $U_s(x)$ in this case depends on 
$x$ rather than on $x_-$. The covariant pure soft building blocks start at 
$\order{\lambda^3}$, for example
\bea
 q(x) \sim \lambda^3,\qquad 
F_s^{\mu\nu} \sim \lambda^4, 
\qquad iD_s^\mu q(x) \sim \lambda^5\,, 
\eea
where {\em on soft building blocks} 
$iD_s^\mu(x)=i\partial^\mu+g_s A_s^\mu(x)$ and the soft field-strength tensor 
is defined as $i g_s F_s^{\mu \nu}= [iD_s^\mu,iD_s^\nu] $. When a pure soft 
building block appears in the product (\ref{eq:Njetop}) with collinear 
fields in several directions, the multipole expansion with respect to each of 
them requires that one sets $x=0$. However, due to the $\order{\lambda^3}$ 
suppression, $J_s(0)$ in Eq.~(\ref{eq:Njetop}) can be dropped at 
$\mathcal{O}(\lambda^2)$. Therefore, soft fields enter neither via the 
soft nor via the collinear building blocks for our basis choice, 
up to ${\cal O}(\lambda^2)$, and option (iii) (fourth panel in 
Fig.~\ref{fig:njetoperators}) is not relevant. 
This implies that the emission of a 
soft gluon from the hard process, which generates the $N$-jet operator, 
is entirely accounted for by Lagrangian interactions. We discuss this 
somewhat surprising result further in the next section. 

The case of $N$-jet operators differs from that of heavy-to-light currents, 
which consist of one collinear direction and a soft heavy-quark field, 
whose decay is the source of large energy for the collinear final state.
The basis of subleading SCET operators listed in 
Ref.~\cite{Beneke:2004in} does contain soft covariant derivatives 
at ${\cal O}(\lambda^2)$ due to the presence of the soft heavy-quark 
building block at leading power. The absence of soft 
building blocks in $N$-jet operators at ${\cal O}(\lambda^2)$ is also 
an important difference and simplification of the position-space 
vs. the label-field SCET formalism \cite{Bauer:2000yr,Bauer:2001yt}, where 
soft fields must be included in the basis operators at 
${\cal O}(\lambda^2)$~\cite{Larkoski:2014bxa,Feige:2017zci}. 
The difference arises from a different split into collinear and soft, 
since in the 
label formalism only the large and transverse component of collinear 
momentum are treated as labels, while the residual spatial dependence 
of all fields, collinear and soft,  is soft. The difference in the 
operator basis due to this is compensated by a corresponding difference 
in the soft-collinear interactions in the Lagrangian in the 
two formulations of SCET.

For the 
derivation of the anomalous dimension and renormalization group equation 
it is convenient to adopt the interaction picture and treat 
the subleading SCET Lagrangians ${\cal L}_i^{(n)}$ as perturbations, 
such that all 
operator matrix elements are understood to be evaluated with the 
leading-power SCET Lagrangian. The basis of subleading power $N$-jet 
operators at a given order in $\lambda$ then includes further 
``non-local'' operators from the time-ordered products of the current 
operators $J$ at lower orders in $\lambda$ with the subleading terms in 
the SCET Lagrangian. The ``local'' (in reality, light-cone) currents do 
not mix into the non-local time-ordered product operators, but the latter 
can, in principle, mix into the former. 
At  ${\cal O}(\lambda)$ the time-ordered product operators are of the 
form 
\be
\label{eq:T1}
J^{T1}_i(t_i) = i \int d^4 x \,
T\left\{J^{A0}_i(t_i), {\mathcal L}_i^{(1)} (x)\right\}\,.
\ee
The generalization to higher orders in 
$\lambda$ should be evident and includes 
subleading currents $J^{A1}$, $J^{B1}, \ldots$, subleading Lagrangians 
${\cal L}_i^{(n)}$ ($n>1$), and time-ordered products with multiple
subleading Lagrangian interactions. In contrast to the local current 
operators, the time-ordered products always contain soft fields. 

For physical applications as well as the presentation of the anomalous 
dimension it is useful to work in terms of collinear momentum fractions by 
defining the  Fourier transforms of the operators with respect to the 
positions $t_{i_k}$ in the collinear direction, 
\be
J_i(P_i,\{x_{i_k}\}) \equiv P_i^{n_i}
\int \prod_{k=1}^{n_i} dt_{i_k}\, 
e^{-i P_i \sum_{k=1}^{n_i} t_{i_k}x_{i_k}}\, 
J_i(\{t_{i_k}\})\,,
\label{eq:Fourier}
\ee
where $x_{i_k}$ are fractions of the collinear momentum in 
direction $i$, carried by the $k$-th building block such that 
$\sum_{k=1}^{n_i} x_{i_k}=1$. $P_i$ is the total 
(outgoing) collinear momentum in direction $i$ and  
$\nnp{i}p_{i_k} = x_{i_k} P_i >0$ 
for an outgoing momentum in direction $i$, such that from 
Eq.~\eqref{eq:Fourier} also $P_i>0$ and $x_{i_k} \in (0,1)$ for all 
momenta outgoing, which we shall assume in the following.
In general, the basis of $N$-jet operators can then be written in the form
\be
  J(\{P_i\},\{x_{i_k}\}) = \prod_{i=1}^N \,J_i(P_i,\{x_{i_k}\})
\ee
The operators are given 
by $J_i\in\{J_i^{An},J_i^{Bn},J_i^{Cn},\ldots\}$, depending on the number of 
collinear building blocks and the order in $\lambda$. 
The total power suppression of the $N$-jet operator is then obtained from 
adding up the suppression factors in $\lambda$ from each direction.
For example, at ${\cal O}(\lambda^2)$, it is possible to either have a 
$J_i^{X2}$ operator (with $X=A,B,C$) in one direction and $J_i^{A0}$ 
operators in the remaining $N-1$ directions, or two operators 
$J_i^{X1}J_j^{Y1}$, with $X,Y=A,B$, and $J_i^{A0}$ operators in the remaining
$N-2$ directions. 


\section{LBK amplitude from SCET} 

\noindent
The Low-Burnett-Kroll \cite{Low:1958sn,Burnett:1967km} formula 
\be
-g_s\sum_{i=1}^N\mathbf{T}_i \left(
\frac{p_i\cdot\epsilon(k)}{p_i\cdot k} + 
\frac{\epsilon_\mu(k) k_\nu J_i^{\mu\nu}}{p_i\cdot k}\right) A_0(\{p_i\})
\label{eq:LBK}
\ee
expresses the radiative amplitude for the emission of a soft gluon 
with momentum $k$ in terms of the corresponding non-radiative amplitude 
$A_0$ (for example, the left panel in Fig.~\ref{fig:njetoperators})
including the NLP correction to the eikonal amplitude. The 
next-to-soft term contains the angular momentum operator 
\be 
J_i^{\mu\nu} = p_i^\mu\frac{\partial}{\partial p_{i\nu}} - 
p_i^\nu\frac{\partial}{\partial p_{i\mu}} +\Sigma_i^{\mu\nu}\,,
\ee
where the spin operator $\Sigma_i^{\mu\nu}=\frac{1}{4}
[\gamma^\mu,\gamma^\nu]$ for a Dirac fermion. The term 
$p_i^\nu\frac{\partial}{\partial p_{i\mu}}$ produces a local contribution 
to (\ref{eq:LBK}), which appears to correspond to an $N$-jet operator 
with a soft building block. In view of the above result that no such 
operators exist at NLP, it is instructive to reproduce the LBK formula 
in position-space SCET.\footnote{The LBK amplitude was analyzed 
in the framework of label-SCET in Ref.~\cite{Larkoski:2014bxa}. It turns 
out that the two formulations of SCET recover the LBK formula in rather 
different ways.}

\begin{figure}[t]
 \centering
 \includegraphics[width=0.55\textwidth]{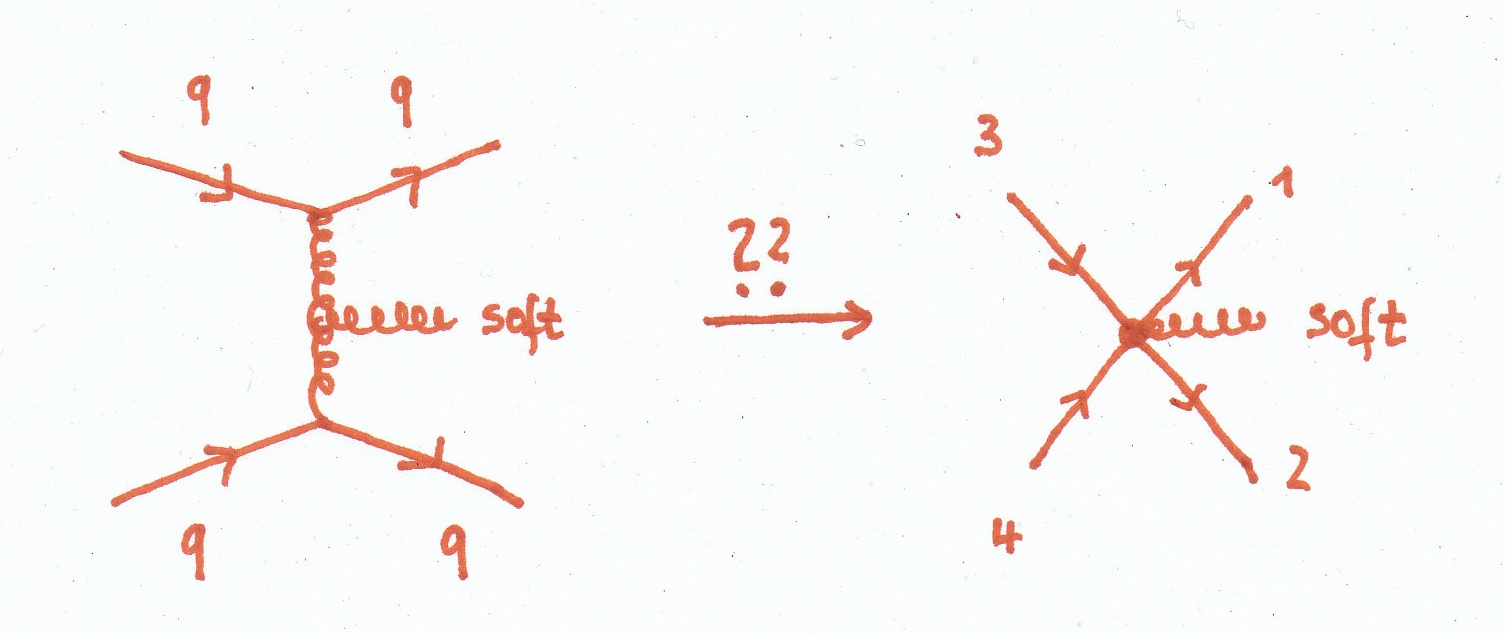}
\caption{Soft emission in quark-quark scattering.
\label{fig:quarkscattering}}
\end{figure}

As an example, we consider the scattering of two quarks of different 
flavour. The tree-level non-radiative amplitude is given by 
\be
A_0 = -g_s^2\,\bar{u}(p_1) \gamma^\rho T^a u(-p_3) \,
\frac{-i}{(p_1+p_3)^2}\,\bar{u}(p_2) \gamma_\rho T^a u(-p_4)\,,
\ee
and carries a non-trivial dependence on the external momenta. The 
tree-level radiative amplitude is given by five diagrams. When the 
soft emitted gluon is attached to the internal propagator as 
shown on the left in Fig.~\ref{fig:quarkscattering}, the duplicated 
internal propagator appears to give rise to the hard four-jet operator 
with a soft gluon building block shown on the right. However, contrary 
to this expectation, we find that the full-theory diagram on the left 
is reproduced to NLP by time-ordered products with the sub-leading 
lagrangian interactions alone, just as the other four diagrams with a 
soft gluon attached to one of the external legs, in the following way:
\begin{itemize}
\item We choose a reference frame in which every $p_i$ is aligned with 
the corresponding collinear reference vector $\nnm{i}$, that is, 
$p_{\perp i}=0$. In this frame the ${\cal O}(\lambda)$ correction 
to the LP eikonal term vanishes, as does the contribution from 
$J^{A2}_i(t_i)$ at NLP ${\cal O}(\lambda^2)$.
\item The time-ordered product 
$\int d^4 x\, T\big\{J^{A0}_i(t_i), {\mathcal L}_i^{(2)} (x)\big\}$ 
with the ${\cal O}(\lambda^2)$ collinear quark-soft gluon interaction 
produces the spin contribution and an additional term, which is 
cancelled by a contribution from $\int d^4 x\, T\big\{J^{A1}_i(t_i), 
{\mathcal L}_i^{(1)} (x)\big\}$. 
\item The orbital angular momentum term also originates fully from 
the above two time-ordered products. The required derivatives on the 
non-radiative amplitude arise from the multipole-expanded soft-gluon 
interaction vertices. For example, $\nnm{i}\cdot x$ in 
${\mathcal L}_i^{(2)} (x)$ yields a factor of $i t_i$ in the convolution 
of the hard coefficient function with the $N$-jet operator along the 
light-like direction, which turns into a momentum-derivative on 
the momentum-space coefficient function/non-radiative amplitude.
\item For this to work the hard coefficient of $J^{A1}_i(t_i)$ must 
be related in a specific way to the one of $J^{A0}_i(t_i)$. The required 
relation follows from invariance of SCET under reparameterizations 
of the light-like direction reference vectors \cite{Manohar:2002fd}.
\end{itemize}
This result appears to be generic, and we checked that it holds as well for 
the heavy-to-light decay considered in Ref.~\cite{Beneke:2004in}. 
Note that, unlike in the original derivation of the LBK amplitude, one 
does not have to invoke gauge invariance and the Ward identity to 
fix the local contribution to the LBK amplitude, 
since gauge invariance is manifest in every 
SCET current and Lagrangian term.

\section{\boldmath 
Anomalous dimension of NLP $N$-jet operators} 

\noindent 
Operator renormalization in renormalized perturbation theory is given by
\bea
\langle {\cal O}_{P}(\{\phi_{\rm ren}\},\{g_{\rm ren}\})\rangle_{\rm ren} 
&=& \sum_Q Z_{PQ}\prod_{\phi\in Q} Z_\phi^{1/2}\prod_{g\in Q} Z_g\langle 
{\cal O}_{Q, \rm bare}(\{\phi_{\rm ren}\},\{g_{\rm ren}\})\rangle\,,
\eea
where $P, Q$ label the $N$-jet operators as well as time-ordered products of 
$N$-jet operators with insertions of power-suppressed interactions 
${\cal L}_{\textrm{SCET}}$. At one-loop, writing 
$Z_{PQ}=\delta_{PQ}+\delta Z_{PQ}$, demanding that the left-hand side is 
finite and accounting for the fact that soft interactions do not 
change the large collinear momentum fractions implies
the $\overline{\rm MS}$ scheme renormalization conditions
\bea
0 &=& \langle J_P(x)\rangle_{\textrm{1-loop, div.}}^{\rm soft, ij} 
+ \sum_{Q}\delta Z_{PQ}^{s, ij}(x) \langle J_Q(x)\rangle_{\rm tree} \,,
\\ 
0 &=& \langle J_P(x) \rangle_{\textrm{1-loop, div.}}^{\rm coll., i} 
+ \sum_{Q}\int \prod_{k>1} dy_{i_k}  \Bigg[ \delta Z_{PQ}^{c, i}(x,y) 
\nn\\
&& {} +\delta_{PQ}\prod_{k>1} \delta(x_{i_k}-y_{i_k})
\left(\frac12 \sum_{\phi\in J_{Pi}} \delta Z_\phi + \sum_{g\in J_{Pi}} 
\delta Z_g\right) \Bigg]\langle J_Q(y)\rangle_{\rm tree} \,
\label{eq:ZABcoll},
\eea
from which the collinear-loop and soft-loop contributions to the 
renormalization factor can be determined. The anomalous dimension matrix is 
defined by
\be
{\Gamma} = - {\bf Z}^{-1} \frac{d}{d\ln\mu} {\bf Z}\,,
\ee
where we use matrix notation involving both discrete indices ($P,Q$) 
labelling the set of $N$-jet operators including open Lorentz, spinor and  
colour indices as well as continuous indices $(x,y)$ for the
collinear momentum fractions associated with each building block. 
To extract the UV divergences, we regularize the IR divergences by 
assuming that the external states have small off-shellness $p_{i_k}^2\neq 0$. 
At the end of the computation, the soft and collinear part are combined 
and only then the limit $p_{i_k}^2\to 0$ can be taken. The cancellation of the 
off-shell regulator dependence serves as an additional check of the 
computation. 

\subsection{\boldmath $F=2$ operators}

In the following, we will focus on the case in which one of the collinear 
directions carries fermion number $F=2$. The simplification of this 
choice results from the absence of a leading-power operator $J_i^{A0}$ 
(and consequently all $J_i^{An}$), since one needs two fermion fields in 
the same direction to begin with. Nevertheless, this simpler case allows 
us to display most of the features of the anomalous dimension at 
${\cal O}(\lambda^2)$. The $F=2$ operator basis at ${\cal O}(\lambda)$ 
consists of the single collinear operator 
\bea
\label{eq:Jxixi}
J^{B1}_{\chi_\alpha \chi_\beta}(t_{i_1},t_{i_2}) &=& 
\chi_{i \alpha} (t_{i_1}\nnp{i})\chi_{i\beta} (t_{i_2}\nnp{i}) \;.
\eea
We keep open the Dirac spinor indices $\alpha,\beta$, because they will in 
general be contracted with components of the $N$-jet operator from the 
other collinear directions $j\not=i$. The same rule applies to Lorentz and 
colour indices, and we only assume that the total $N$-jet operator transforms 
as a colour singlet. The ${\cal O}(\lambda^2)$ basis operators are 
constructed as described in Sec.~\ref{sec:basis} and include the 
three-body operator
\bea
J^{C2}_{{\cal A}^\mu \chi_\alpha \chi_\beta}(t_{i_1},t_{i_2},t_{i_3}) &=&  
{\cal A}_{\perp i}^\mu(t_{i_1}\nnp{i}) 
\chi_{i\alpha}(t_{i_2}\nnp{i})\chi_{i\beta}(t_{i_3}\nnp{i})\,.
\label{eq:Jxixi2}
\eea
We refer to Ref.~\cite{Beneke:2017ztn} for the details of this 
computation and restrict ourselves here to the following remarks:
%
\begin{figure}[t]
 \centering
 \includegraphics[width=0.23\textwidth]{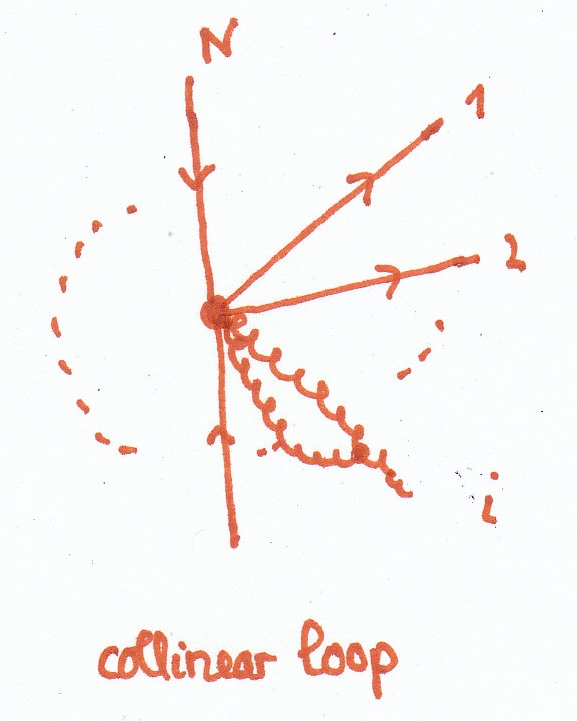}\hspace{0.8cm}
 \includegraphics[width=0.22\textwidth]{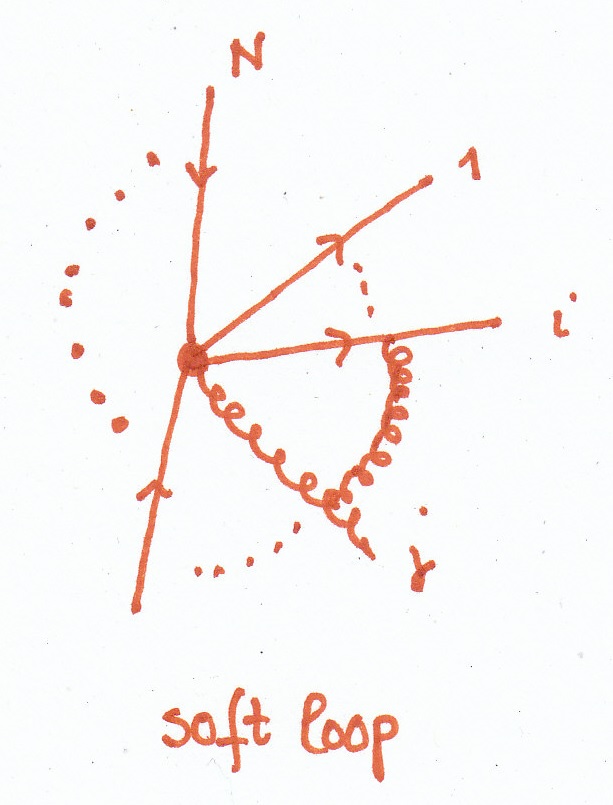}
 \caption{Collinear (left) and soft (right) loop diagram.
\label{fig:scloops}}
\end{figure}
%
\begin{itemize}
\item Collinear loops always connect lines within the same collinear 
sector, while soft loops always connect lines of two different collinear 
directions, see Fig.~\ref{fig:scloops}.
\item Collinear loops within a single building block (as in the left 
panel of Fig.~\ref{fig:scloops}) are trivial. However, collinear 
loops connecting 
different building blocks in the same collinear direction change the 
distribution of momentum fraction and result in momentum-dependent 
renormalization factors. For example, the collinear contribution to the 
anomalous dimension of the B1 operator (\ref{eq:Jxixi}) reads
\bea\label{eq:ZJxixi}
\delta Z^{c,i}_{\chi_{\alpha}\chi_{\beta},\chi_{\gamma}\chi_{\delta}}(x,y) 
&=& - \delta(x-y)\delta_{\alpha \gamma}\delta_{\beta \delta} X_{i_1i_2} + 
\frac{1}{\epsilon}\,\gamma^i_{\chi_{\alpha}\chi_{\beta},\chi_{\gamma}
\chi_{\delta}}(x,y)\,,
\eea
with
\bea
  X_{i_1i_2} &\equiv& \frac{\als}{4\pi} \Bigg\{\frac{2}{\epsilon^2}({\bf T}_{i_1}+{\bf T}_{i_2})^2 + \frac{2}{\epsilon}({\bf T}_{i_1}+{\bf T}_{i_2})\cdot\Bigg[{\bf T}_{i_1}\ln\left(\frac{\mu^2 }{-p_1^2}\right) \nn\\
  && + {\bf T}_{i_2}\ln\left(\frac{\mu^2 }{-p_2^2}\right) \Bigg] + 
\frac{3}{2\epsilon}\left({\bf T}_{i_1}^2+{\bf T}_{i_2}^2\right)  \Bigg\} \,,
\eea
and
\bea\label{eq:gammaxixi}
\gamma^i_{\chi_{\alpha}\chi_{\beta},\chi_{\gamma}\chi_{\delta}}(x,y) &=& \frac{\als {\bf T}_{i_1}\cdot{\bf T}_{i_2}}{2\pi} \Bigg\{ \delta_{\alpha\gamma}\delta_{\beta\delta}   \Bigg( \theta(x-y)\left[\frac{1}{x-y}\right]_+ + \theta(y-x)\left[\frac{1}{y-x}\right]_+ \nn\\
  &&  {} - \theta(x-y)\frac{1-\frac{\bar x}{2}}{\bar y} -\theta(y-x)\frac{1-\frac{x}{2}}{y}\Bigg)\nn\\ 
  && -\frac14 \left(\sigma_\perp^{\nu\mu} \right)_{\alpha\gamma}\left(\sigma_{\perp\nu\mu}\right)_{\beta\delta}  \left(\theta(x-y)\frac{\bar x}{\bar y}+\theta(y-x)\frac{x}{y}\right) \Bigg\}\,.
\eea
\item At ${\cal O}(\lambda^2)$ there is operator mixing among the 
B2 and C2 operators, but C2 operators do not mix into B2 operators.
\item The renormalization factor of the three-body operator 
(\ref{eq:Jxixi2}) follows from the ones of two-body quark-quark 
and quark-gluon operators, since in the one-loop order only two collinear 
building blocks can be connected through a loop.
\item For the $F=2$ case considered, the soft renormalization is trivial. 
It is diagonal in operator space and momentum fraction, spin-independent, 
and has the same dipole form as the first term in 
Eq.~(\ref{eq:LPanomalousdim}). The possible mixing of the time-ordered 
product operators into the current operators also vanishes.
\end{itemize}

\subsection{Structure of the anomalous dimension}
\label{sec:result}

\noindent
Summing over all $N$ collinear directions, combining the soft 
and collinear loop contributions, and further using colour conservation 
for the total (colour-singlet) $N$-jet operator, we can cast the 
one-loop anomalous dimension into the form 
\bea\label{eq:main}
\Gamma_{PQ}(x,y) &=& \delta_{PQ}\delta(x-y)\!\left[ 
-\gamma_{\textrm{cusp}}(\als)\sum_{i<j}
\sum_{l,k}{\bf T}_{i_l}\cdot{\bf T}_{j_k}\hspace{-0.02cm}
\ln\left(\frac{-s_{ij}x_{i_l}x_{j_k}}{\mu^2}\right) 
+ \sum_i\sum_l \gamma_{i_l}(\alpha_s) \right] 
\nn\\
  && + \,2\sum_i\delta^{[i]}(x-y)\gamma^i_{PQ}(x,y)\,,
\eea
where $\gamma_{\textrm{cusp}}(\als) = \frac{\als}{\pi}$, 
$\gamma_{i_l}(\alpha_s)\equiv -\frac{\als}{2\pi}{\bf T}_{i_l}^2 c_{i_l}=-\frac{3\als}{4\pi}C_F \, (0)$ for collinear quarks (gluons). The short-hand  
delta-functions are defined as 
 $\delta(x-y)\equiv \prod_{i}\prod_{k=2}^{n_i}\delta(x_{i_k}-y_{i_k})$,
 $\delta^{[i]}(x-y)\equiv 
\prod_{j\not=i}\prod_{k>1}\delta(x_{j_k}-y_{j_k})$.
The second line captures the off-diagonal contributions such as 
Eq.~(\ref{eq:gammaxixi}) and arises only from single $1/\epsilon$ poles. 
The first line is diagonal and contains a logarithm whose coefficient 
is related to the familiar cusp anomalous dimension.

We note the similarity of the first line with the leading-power 
anomalous dimension~(\ref{eq:LPanomalousdim}). There is an additional 
sum over the number of building blocks in every collinear direction. 
Indeed, Eq.~\eqref{eq:main} reduces to  
Eq.~(\ref{eq:LPanomalousdim}) when there is only a single building
block in each collinear direction (i.e. $l,k= 1$, $x_{i_l},x_{j_k}\to 1$), 
such that in the notation used above $\delta(x-y)\equiv\prod_i\prod_{k>1}\delta(x_{i_k}-y_{i_k})\to 1$ is an empty product equal to unity. Furthermore, 
possibly non-diagonal contributions encapsulated in $\gamma^i_{PQ}$
vanish at the leading power.

For the case in which one of the collinear directions contains two fermionic 
building blocks (direction $i$, say), there is only a single $N$-jet operator 
of this kind at ${\cal O}(\lambda)$, given by the product of $
J_i=J^{B1}_{\chi\chi}(t_{i_1},t_{i_2})$ defined in Eq.\,\eqref{eq:Jxixi}
for the direction labelled by $i$ and leading-power building blocks for all 
other $N-1$ directions $J_{j\not=i}=J_j^{A0}$. In this case, the anomalous 
dimension is off-diagonal in the collinear momentum fractions in direction $i$,
\be
\sum_{j=1}^N\delta^{[j]}(x-y)\frac{\gamma^j_{PQ}(x,y)}{\epsilon} 
\to \frac{1}{\epsilon}\,\gamma^i_{\chi\chi,\chi\chi}(x_{i_1},y_{i_1})\,,
\ee
where the right-hand side is given by Eq.~\eqref{eq:gammaxixi}, and we have 
used $\gamma^j_{PQ}(x,y)= 0$ for all leading-power building
blocks $j\not= i$. Furthermore the product of delta functions for the $N-1$ other directions 
$\delta^{[i]}(x-y)\equiv\prod_{j\not=i}\prod_{k>1}\delta(x_{j_k}-y_{j_k})\to 1$
also collapses to unity.

At ${\cal O}(\lambda^2)$, there are two possibilities. First, 
that the direction $i$ which we choose to carry fermion-number two 
encompasses itself the ${\cal O}(\lambda^2)$ suppression, i.e. it is 
represented by one of the three operators 
$J_i\in \{J^{B2}_{\chi\partial\chi},J^{B2}_{\partial(\chi\chi)},
J^{C2}_{{\cal A}\chi\chi}\}$.
Then the other $N-1$ directions have to contain leading-power building 
blocks, as before. There is operator mixing among the three 
${\cal O}(\lambda^2)$ suppressed operators and the  structure of the 
anomalous dimension matrix is 
\be
  \sum_j\delta^{[j]}(x-y)\frac{\gamma^j_{PQ}(x,y)}{\epsilon} \to \frac{1}{\epsilon}
  \left(\begin{array}{ccc}
  \gamma^i_{\chi\partial\chi,\chi\partial\chi} & \gamma^i_{\chi\partial\chi,\partial(\chi\chi)} & \gamma^i_{\chi\partial\chi,{\cal A}\chi\chi} \\
  0 & \gamma^i_{\partial(\chi\chi),\partial(\chi\chi)} & 0 \\
  0 & 0 & \gamma^i_{{\cal A}\chi\chi,{\cal A}\chi\chi} 
  \end{array}\right)\,,
\ee
where the non-zero contributions are given in Ref.~\cite{Beneke:2017ztn}. 
The anomalous dimension is diagonal with respect to the other $N-1$ directions.

The second possibility that can occur at ${\cal O}(\lambda^2)$ is 
that direction $i$ 
with $F=2$ is described by the ${\cal O}(\lambda)$ contribution 
$J_i=J^{B1}_{\chi\chi}(t_{i_1},t_{i_2})$, and one of the other $N-1$ 
directions, say direction $i'$, contributes an additional ${\cal O}(\lambda)$ 
suppression. The remaining $N-2$ directions must then be represented by 
leading-power building blocks. Since we do not require direction $i'$ to have
a definite fermion number, there are more possibilities, in particular $J_{i'}\in\{J^{A1}_{\partial\chi}, J^{A1}_{\partial{\cal A}}, J^{B1}_{{\cal A}\chi},
J^{B1}_{\cal AA}, J^{B1}_{\chi\chi}, J^{B1}_{\bar\chi \chi}\}$ (plus hermitian conjugated operators). In this case, we need in addition the corresponding 
anomalous dimension matrices $\gamma^{i'}_{PQ}$ for these operators.

\section{Summary}

\noindent 
The anomalous dimension of subleading power $N$-jet operators is one of 
the key ingredients for the resummation of logarithmically enhanced terms 
in partonic cross sections beyond the leading power. Schematically, 
an observable expanded in powers of a small scale-ratio $M/Q$ 
will be represented in the form 
\be
d\sigma = \sum_{a,b} C_a C_b^*\,
\otimes \,J_{ab} \otimes \,S_{ab}
\ee 
of a convolution of the product $C_aC_b^*$ of 
hard matching coefficients, jet and soft functions. 
The $N$-jet operator anomalous dimension discussed here 
governs the renormalization-group evolution of hard 
functions of all possible operators up to NLP, which are convoluted with 
their respective jet and soft functions. 
For a given process the colour- and spin-space 
anomalous dimensions have to be projected on a basis of scalar 
operators. The renormalization-group equations are integro-differential 
equations in every collinear sector, which will most likely have to 
be solved approximately or numerically. 

\subsubsection*{Acknowledgements} 
\noindent 
This work has been supported by the BMBF grant no. 05H15WOCAA.

\bibliography{procsNLP}

\end{document}